\documentclass[11pt]{article}
\usepackage{rotating}
\usepackage{cite}

\begin{document}
\pagenumbering{roman}
\begin{titlepage}
 
\begin{center}
{\large A Case Study of Gender Bias at the Postdoctoral Level in Physics,\\ and
its Resulting Impact on the Academic Career Advancement of Females}
\end{center}

\begin{center}
   {\large
          Sherry Towers BSc MSc PhD\\[2mm]
   }
   {\small
          Purdue University\\
          Department of Statistics\\
          West Lafayette, IN\\
          47907-2066\\[3mm]
          {\tt stowers@purdue.edu}\\
          
   }
\end{center}
 
\begin{center}
{\large {\bf Abstract}}
\end{center}

This case study 
of a typical U.S. particle physics experiment 
explores the issues of gender bias and how it affects the academic career advancement 
prospects of women in the
field of physics beyond the postdoctoral level; 
we use public databases to study the career
paths of the full cohort of 
57
former postdoctoral researchers on the Run~II~Dzero experiment
to examine if males and females were treated in a gender-blind fashion on the
experiment.  

The study finds that the female researchers 
were on average significantly more productive compared to their male peers, 
yet were allocated only 1/3 the amount of conference presentations based
on their productivity.  The study also finds 
that the dramatic gender bias in allocation of 
conference presentations appeared to have
significant negative impact on the academic career advancement of the females.

The author has a PhD in particle physics and worked for six years as a postdoctoral
research scientist, five of which were spent collaborating at Fermilab.  
She is currently completing a graduate degree in statistics.

\end{titlepage}
\pagenumbering{arabic}

\section{Introduction}

35 years ago, when the U.S. academic anti-discrimination law Title IX was first
enacted\footnote{See {\tt http://www.usdoj.gov/crt/cor/coord/titleixstat.htm} for 
a description of the law.}, women with doctoral degrees in the fields of science
rarely made it to the faculty level, and physics was the worst of these fields
in this respect. 
Over the past few decades some advances have been made in all fields of
science in the increase in the fraction of women at the faculty level. 
Today however physics
still ranks as the worst of the fields of science in this respect (see reference Nelson (2005)),
and it is clear that Title IX has so far had little impact on dramatically increasing the
participation and retention of women in physics, unlike the impact it has had
on other fields such as athletics (see references USGAO (2004) and Rolison (2003)). 

A law such as Title~IX can improve the retention of women in physics only if the
reasons for gender inequities in academic career advancement are primarily due
to gender discrimination in the field rather than free career path choices made by
physicists. For example, as hypothesized in 2005 
by {Dr.Larry Summers}, then president of Harvard, if women 
innately prefer non-scientific fields and/or are innately incapable of performing scientific
research at the standards required at the higher echelons of the field, there is
nothing an anti-discrimination law can do to retain them in the field.        
As has been frequently noted in the hot debate surrounding Summers' comments,
disentangling the reasons for the continuing gender inequities 
in scientific academic career advancement 
is a complex process, 
especially when one tries to determine
the factors that appear to have the
most influence on career paths, and whether or not
those factors are gender dependent and/or within the control of the scientist.  

Relying upon survey data for such studies unfortunately carries the dangers of survey bias.
Ideally a data set is needed that allows for unbiased assessment of a scientist's productivity,
career advancement perks awarded to the scientist, and the
eventual career path of the scientist.  Even better is such a data set that follows
a full cohort of scientists of about the same age who work
under the same power infrastructure during the same period of time. 

Just such an analysis was performed 10 years ago by
Wenneras and Wold (1997). In their landmark
study they 
investigated whether the Swedish Medical Research Council (MRC), 
one of the main funding agencies for biomedical research in Sweden, 
evaluates women and men on an equal basis.
Their study examined the productivity of the applicants, and
showed that female applicants for postdoctoral fellowships were strongly 
disfavored over men with the same productivity. 
Wenneras and Wold also examined the impact of ``socialization'' on the academic
career advancement prospects of females, and they found that females who were socially
connected to one or more members of the MRC were much more likely to be awarded
a postdoctoral fellowship.

This study is similar in many respects to that of Wenneras and Wold, except 
that we examine a cohort of postdoctoral researchers in particle physics rather than biomedicine;
in our study we examine  
a richly detailed set of public online databases maintained by the Run~II~Dzero experiment\footnote{See 
{\tt http://www-d0.fnal.gov} for a description of the experiment.},
a typical U.S. particle physics experiment in both its size and infrastructure.
These databases allow any member of the general
public to delve surprisingly deeply into the daily workings of the experiment.
For instance, the Run~II~Dzero experiment maintains a public searchable
online comprehensive database
that keeps track of internal Run~II~Dzero
papers; these papers document the ongoing work of
each researcher on the experiment, and the number of such papers each
researcher writes is an excellent measure of his or her productivity since all collaborators
on the experiment are expected to document their work in a timely fashion.
In addition, the experiment maintains an online public database
that keeps track of which Run~II~Dzero collaborators are allotted
presentations at professional physics conferences.  Conference presentations
provide young researchers much needed exposure to potential future
employers, yet collaborators on the experiment can give
a conference presentation {\bf only} if they are allowed to do so by the upper administration of the experiment.

In our study we follow the career paths of the full cohort of postdoctoral
researchers who collaborated on the Run~II~Dzero experiment between 1998 and the end of 2006, and who have since moved on
to another job.
In this study, similar to Wenneras and Wold, we assess the productivity of each researcher in our sample,
and then examine whether or not the reward (in this case conference presentations, in the case of Wenneras and Wold,
postdoctoral research grants) is gender-blind.  
We also determine whether or not the number of physics conference
presentations allocated to a female researcher is associated with her prospects of academic career
advancement.

In addition, similar to Wenneras and Wold, we assess the ``socialization'' of each researcher in our
sample, in this case through examination of the average number of co-authors with whom each researcher publishes.
We determine whether or not the 
socialization of a 
female researcher is associated with her prospects of academic career
advancement.

Our cohort of past researchers on the Run~II~Dzero experiment consists of
48 males and 9 females.
Before we continue, it is worthwhile to briefly discuss the issue of 
small sample sizes and whether or not information regarding discriminatory bias can be 
gleaned from this sample.
In order to make definitive statements about differences between males and females on the basis 
of a relatively small sample of females 
we use the 48 males as a reference
to which we can compare the properties of a particular female.
As we will describe in the analysis section of this paper, we can form a robust
statistical test to determine if a property of a particular female (such as productivity)
is less than or greater than that of a typical male.   Thus, rather than relying solely upon
averages of data to compare males to females (a procedure which is somewhat
limited by our small sample of females), we will instead predominantly be using these statistical
comparison tests.

To examine the properties of males and females who move on to faculty positions compared to those
who do not, we develop separate linear regression models for the males and females that predict the probability of
becoming a professor with a linear combination of the measures of researcher productivity, socialization, and
conference reward ratios.  We trim each model to obtain the most parsimonious model that best describes each
group, and determine the model variables that appear to be most significant.
We then determine whether or not the variables that best predict academic career advancement are gender dependent.

In the following sections we will describe the Run~II~Dzero
experiment and its administrative
infrastructure, followed by a description of the selection of the cohort of postdoctoral researchers
used in this study, and a description of the analysis of the data.

\section{Run~II~Dzero: A Typical U.S. Particle Physics Experiment}
Run~II of the Dzero experiment is a 
multi-institutional
international collaboration of almost 700 experimental particle physicists, based 
at the Fermi National
Accelerator Laboratory (Fermilab) near Chicago, Illinois.
This experiment is a typical U.S. particle physics experiment in both its size and 
administrative infrastructure.

Run~II of the Dzero experiment began taking data in March 2001 and
is expected to continue taking data until sometime approximately between 2008 and 2010.
The physics results produced by the Dzero experiment are quite diverse, and there
are expected to be over 100 papers associated with these results
published in refereed journals
during Run~II of the experiment.

The Run~II~Dzero experiment is a sister experiment to the Run~II~CDF experiment, 
which is also based at Fermilab\footnote{See {\tt http://www-cdf.fnal.gov} for 
a description of the experiment}.  
In many respects, the two experiments are essentially
in competition to be the first to produce various physics results.
Because of the keen competitive nature of the two experiments,
a need is felt to disseminate significant results in very short order.
Thus physics results produced by the experiments are almost always 
first disseminated to the outside world via one or more
presentations given at professional physics conferences (which happen
regularly throughout the year), followed later by publication in refereed journals. 

Collaborators on the Run~II~Dzero experiment may give conference presentations 
only if they are given permission to do so by the upper 
administration of the experiment. This autocratic practice is typical among
particle physics experiments in America, although the exact details of how conference
presentations are
allocated vary from experiment to experiment.

Conference presentations are  
important to the career advancement prospects of postdoctoral 
particle physicists 
primarily because they give a young
physicist much needed positive exposure to future employers (important because
postdoctoral positions are inherently temporary).
Conference presentations are also important to career advancement
because the names of all of the several hundred
collaborators are included on the author list of
refereed publications. The names appear in alphabetical order only, and there are no first authors.
This is because particle physics in theory strives to be an egalitarian field where the contributions
of all physicists in the collaboration are given equal merit.

Unfortunately, this results in a system where physicists outside the experiment
have difficulty in determining which physicists were primary participants in
a Run~II~Dzero analysis.
Conference presentations provide one of the only means available to disentangle this;
the allocation of a conference presentation on the topic of a particular analysis
states to the outside world that the presentor was ostensibly a primary participant in the work,
and that the upper administration of the experiment considers
the presentor to be an active and productive collaborator.
 
At Run~II~Dzero, conference presentations are allocated
by a panel of 6 to 8 people who are appointed 
by the upper administration of the experiment.
The minutes of the meetings in which the conference presentation
allocations are decided are not made public to the collaborators
on the experiment.  It is a system that may unfortunately be prone to
potential cronyism and patronage; both in who gets appointed to the panel,
and to whom the panel allocates conference presentations.  

Work done by researchers collaborating on the experiment is divided into physics
and ``service work''. 
Service work includes tasks such as detector
development and operation, calibration and alignment of the detector, and other tasks such
as algorithm development for particle recognition.  Service work must necessarily be
performed to ensure that the experiment runs smoothly.  Ostensibly, all collaborators
on the experiment are expected to contribute to service work for at least 12 hours per week.  
For junior physicists,
such as postdoctoral researchers, the type of service work they perform (and the amount of
service work they do) is usually not their decision.  This is instead dictated by
the upper
administration of the experiment and/or their employer.  Service work is generally viewed as a necessary
evil by those collaborating on the experiment, since such work almost never in and of itself
produces publications in refereed journals.  Service work is thus not usually
amenable to the academic career
advancement of junior physicists, because these researchers 
typically must show that they are capable of co-ordinating
a comprehensive independent {\it physics} research program in order to progress.
In fact, as we will see later, the more service work a researcher performs, the more it negatively impacts
their probability of academic career advancement.

In contrast, work done on physics generally results in publications in refereed
journals, and indeed such publications are the {\it raison d'\^{e}tre} of the experiment.
Physics analysis work is normally much more
amenable to the career advancement of a postdoctoral researcher since
the resulting journal publications associated with their
work can be listed on their vita, and such work shows that the postdoctoral
researcher would likely be capable of overseeing an independent physics research
program as a faculty member.

Junior physicists are sometimes given freedom to choose which physics analyses they work
upon, but often subtle or overt influence is put upon them by the administration of the
experiment and/or their employer to work on specific physics analyses.
Junior physicists who are well mentored and/or well connected on the experiment often
are able to work on ``hot'' (high-profile) physics analyses. Competition to work on these
high-profile analyses is usually intense and one of the few ways a junior physicist can
become involved in such analyses is via help from a mentor who is actively engaged in
promoting the career advancement of their prot\'{e}g\'{e}(e).

A full version of the Run~II~Dzero governance document, including information on
the experiment's policy for the allocation of conference presentations and service work requirements,
can be found at  
{{\tt http://d0server1.fnal.gov/projects/Spokes/documents/d0\_manage\_oct00.html}}
As of March 2007 this page was publicly accessible.

\section{Data Selection}

Since 2002 the Run~II~Dzero experiment has maintained, on an
approximately annual basis, lists of postdoctoral researchers then currently
serving on the Run~II~Dzero experiment\footnote{An explanation of why data
on the career paths of postdoctoral researchers is recorded can be found in Appendix VIII, section C.4 of the Run~II~Dzero governance
document:\\ {{\tt http://d0server1.fnal.gov/projects/Spokes/documents/d0\_manage\_oct00.html}}\\
An example of such data can be found at {{\tt http://www-d0.fnal.gov/ib/oct03}}\\
(as of March 2007 these pages were publicly accessible).
}.
The lists include the name of the institute employing the researcher, along with
their start date.  These lists show that Run~II~Dzero postdoctoral researchers
were hired almost exclusively after 1998. 
The Run~II~Dzero experiment also maintains a list of past
postdoctoral researchers that includes all of the above information,
along with the researcher end date and job position obtained after the
completion of the postdoctoral term.

Using these combined sources of information, along with the changing membership
of past Dzero author lists on published papers,
supplemented with information publicly available on the internet,
we compiled a list of past postdoctoral researchers on the Run~II~Dzero experiment.
Information gleaned from the internet included, amongst other things, vitae and/or current
employment of past Dzero postdoctoral researchers, and
lists of researchers previously employed by Dzero institutions. 
The author made every effort using the above data to ensure that the resulting list of
past Run~II~Dzero postdoctoral researchers is as comprehensive as possible.

In this study we focus on postdoctoral researchers who were hired on or after 1998,
served as researchers on the Run~II~Dzero experiment
for at least two years (such that they would appear on
the Dzero author list, and also have a significant period of time
invested in that position), and who performed research only (or almost exclusively) on the
Run~II~Dzero experiment during the tenure of their research position (such that their career
path would be influenced primarily by their work on Run~II~Dzero, and not on some other experiment). 
Additionally, we only examine
Caucasian researchers in this study to avoid any confounding of our results
due to potential additional racial bias.

Collection of data for this study was completed in early 2007.  Thus only postdoctoral
researchers who had moved on to a different job by the end of 2006 are considered in
this study.  

In order to ensure that a researcher's presence and participation in
collaboration at the laboratory is not limited
by geographic constraints (requiring international travel, for instance),
we require all researchers in the sample to be employed by
U.S. institutions.
 
It should be noted that nearly all the postdoctoral researchers included in this
study were employed by U.S. universities rather than by Fermilab; these researchers were based
at the laboratory to perform research there, but were not actually paid by the laboratory
itself.  Yet, as described above, many aspects of their day-to-day working lives were controlled
by the administrative infrastructure of the experiment rather than the administrative infrastructure of
the universities that employed them.

The final sample of past postdoctoral researchers consists of 48 males and 9 females. The
years-spent-in-postdoctoral-position distribution of the two samples is statistically
similar (males and females spend an average of $3.9\pm0.2$ and $4.2\pm0.5$ years in their position, respectively).
16 out of the 48 males, and 4 out of the 9 females went on to faculty positions, respectively.

\section{Analysis}
The following sections 
examine gender differences in researcher productivity, conference rewards, and socialization. We
then build linear models based on these variables to predict the academic career outcomes of males and females.

\subsection{Researcher Productivity}

The Run~II~Dzero experiment maintains a searchable
online comprehensive database
that keeps track of internal Run~II~Dzero
papers; these papers document the ongoing work of
each researcher on the experiment\footnote{See {\tt http://www-d0.fnal.gov/d0notes\_forms/d0noteSelMin.html} 
(as of March 2007 this was a publicly accessible database)}.
All researchers on the Run~II~Dzero experiment are expected to document
their work in internal papers on a regular basis, such that their
work can be collaboratively shared with the rest of the people participating
on the experiment.  
We thus use the number of internal papers authored or co-authored by each
researcher in our sample as a measure of their productivity.
We divide the internal papers into papers documenting physics analyses,
and those documenting service work
to the collaboration.

Interesting differences between the productivity of the males and females 
in our sample our
seen;  24 of the males (exactly half the sample of males) produced fewer internal papers per year than
the {\it least} productive female in the sample.  The higher productivity of the females appears
to translate into somewhat greater chances of career advancement; 16 (33\%) of the 48 males went on
to become faculty members, whereas 4 (44\%) of the 9 females became faculty members (however, it should be noted
that the difference between these two fractions is not statistically significant).
Of the 24 males who were at least as productive as the least productive female, 
11 (46\%) became faculty members.
This similarity between the fractions of females and ``equivalently productive'' males who become faculty
appears at first blush to be a sign of equity in promotion and hiring\footnote{Assuming of course that equal fractions of females and males
want to become faculty members, which is likely a good assumption given the academic career level
already reached by the members of the cohort. If we assume that high productivity results from a desire
to progress up the academic career ladder, it may even be concluded that females in this cohort appear to desire
academic career advancement more than their male peers.}, where hard work is rewarded with career advancement.
However, as we will see below, the productivity of females is significantly higher even than that of
the typical male in this somewhat mis-named ``equivalently productive'' sample of males.

As an aside, we note that 20\% of the 24 males who produced almost nothing nevertheless
went on to faculty positions, and that 19/24 (79\%)
of these males
were allocated at least one conference presentation, whereas only 6/9 (66\%) of the females were allocated conference
presentations despite their very high productivity.  In fact, the mean number of presentations per year awarded to these
males slightly exceed the mean number awarded to the females, and the conference reward ratio for the
unproductive group of males is over four times greater than that of the females ($p<0.1\%$).

Table~\ref{tab:comp} displays the averages of measures of productivity for the full sample of males and females.
Also shown is the standard error on each average, and
in brackets by each average is the correlation of each variable to the probability that a female or male becomes a faculty member.
We see that the probability that a male advances up the academic career ladder is significantly correlated to his total and physics
productivity 
For females, the probability that they will become a professor is significantly anti-correlated to the fraction of
their productivity devoted to service work.  

It has been mentioned above that the nature of a particle physics collaboration requires that all collaborators document their
work in a timely fashion such that it can be shared with their collaborators.  However, if there
in fact exists some type of work that does not necessarily produce documentation in the form of internal papers,
that type of work is presumably performed equally by females and males (under the assumption of no gender bias).
Thus it cannot be argued that internal papers do not form a measure of productivity; internal papers may not
be an {\it absolute} measure of productivity, but they give an equitable assessment of how
much relative work males and females are performing in the collaborative environment.

In the last column of the table we show "one-sided significance test probabilities" that use
the 48 males as a reference sample to test whether a particular
variable for females is significantly less than (or greater than) that of the typical male.  
Using researcher productivity as an example,
we first examine the productivity of a particular female and
determine how many males have productivity less than hers.
If a particular female is much more productive than most males, the fraction of males with 
productivity less than hers will be close to 1.  
Because the sample size of 48 males is quite large,
this is a robust statistical test (see references
Epstein (1954), and Coberly and Lewis (1972))
We will call this probability an ``upper-side'' significance test probability and denote it by $P_{\rm female>males}$.
We interpret this as a probability that
tests the hypothesis that a particular variable
for a female was randomly drawn from the corresponding reference distribution for the males.
If females are the same as males, this probability will be uniformly distributed between 0 and 1, but
if females are on average quite different than males for some variable, 
the corresponding upper-side test probability will
either be close to 0 or close to 1.

To combine the individual probabilities for each of the 9 females to obtain the overall $P_{\rm females>males}$ for
the full sample of females,
we begin by multiplying all of these probabilities for the individual females to obtain
$A = \prod_{i=1}^{N} (P_{\rm female>males})_i$, where $N$ is the number of females.
The overall $P_{\rm females>males}$ probability is then (see reference Fisher (1948))
\begin{eqnarray}
P_{\rm females>males} = A \sum_{i=0}^{N-1} {{(-\log{A})^i}\over{i!}}.
\nonumber
\end{eqnarray}
Under the null hypothesis that the males and females have the same underlying probability
distribution for that variable, the $P_{\rm females>males}$ value will be uniformly distributed between 0 and 1.
In our example, if the overall upper-side test probability for the productivity of the female
sample is very close to 1, we can conclude that the cohort of females appears to have  
significantly higher productivity than that of the typical male.  

Note that sometimes the means between the males and females may not differ significantly for some variable, but
$P_{\rm females>males}$ is significant (say, larger than $0.90$).  This merely indicates that males and females
have differently shaped data distributions for that variable, but those distributions happen to have similar means.
Thus it can be seen that the upper-side test probability provides more information than just
a simple comparison of averages between groups because it includes information about
possible differences in the shapes of the distributions of the two groups.

Similar to $P_{\rm females>males}$,  we can define a ``lower-side'' significance test probability $P_{\rm females<males}$.
For instance, if a particular female has a much smaller conference reward ratio than most males, the fraction of males with 
conference reward ratios greater than hers will be close to 1.  
We denote this lower-side test probability by $P_{\rm female<males}$, and
again we combine the lower-side test probabilities for each of the females to
form an overall lower-side test probability, $P_{\rm females<males}$ for the sample of 9 females.

For large reference samples, $P_{\rm females>males}$ approaches $1-P_{\rm females<males}$.
In this analysis we quote the maximum of the upper- and lower-side probabilities, 
because statistical complications arise in the assessment of a one-sided test probability for the full sample of females
when the one-sided test probability for at least one of the females is zero.
These statistical complications are beyond the scope of this paper.

From the one-side test probabilities in the Table~\ref{tab:comp} we see that females appear to be 
significantly more productive than males for all measures of productivity, and perform
significantly more service work (on average around 40\% more service work than males).
As an aside it should also be noted that even if we compare the productivity of the females to that
of the ``equivalently productive males'' (ie; males with productivity at least as large as that
of the least productive female) we find that $P_{\rm females>males}=0.95$.  Thus even these ``equivalently
productive'' males do not on average meet the high productivity standards of the females.

We note from the one-sided tests that the females are significantly more productive than males for all productivity measures, even though the
averages of some productivity measures (such as the number of physics internal papers produced per year)
appear to be somewhat similar; this is due to the different shapes of the productivity distributions of the males and females;
nearly all the females are highly productive, whereas 1/2 the males produce almost
nothing, somewhat less than half are moderately productive, and a select few are extremely productive.
Since the females
have a greater physics productivity than the bulk of the males, the upper-sided test probability is high despite
the similarity in the means of the physics productivity distributions.

\subsection{Allocated Conference Presentations}
The Run~II~Dzero experiment maintains a public searchable
online comprehensive database
that keeps track of which Run~II~Dzero collaborators are allotted
presentations at particular conferences\footnote{See {\tt http://d0trigdb01.fnal.gov:8080/D0speakers/www/index.html}
(as of March 2007 this was a publicly accessible database).
}.
In this study we use this database to determine how many conference
presentations were allocated to Run~II~Dzero postdoctoral researchers in our sample.
We divide the conference presentations into those devoted to physics only, and those devoted
to service work topics.

As previously mentioned, conference presentations
are important to the career advancement prospects of postdoctoral 
particle physicists because 
giving a presentation at a large physics conference gives a young
physicist much needed exposure to future employers (important because
postdoctoral positions are inherently temporary), and they also announce
to the outside world (either rightly or wrongly) that the researcher played
a significant role in the analysis being presented.

Since conference presentations are ostensibly allocated to those who play
roles in the completion of an analysis, we would expect that
the number of conference presentations allocated is directly related to the
productivity of the researcher.  
We thus define a conference presentation ``reward ratio'' as
\begin{eqnarray}
{\rm total\hspace*{0.2cm}conference\hspace*{0.2cm} reward\hspace*{0.2cm} ratio} = 
{{\rm \#\hspace*{0.2cm} allocated\hspace*{0.2cm} conference\hspace*{0.2cm} presentations}\over
{{\rm total\hspace*{0.2cm} \#\hspace*{0.2cm} of\hspace*{0.2cm} internal\hspace*{0.2cm} papers\hspace*{0.2cm} produced}+1}}.
\end{eqnarray}
The addition of 1 in the denominator is needed to ensure that the reward ratio is defined for
people who are awarded conference presentations despite zero productivity (which occurs for some males).

In a similar fashion we can define a physics conference presentation reward ratio as
\begin{eqnarray}
{\rm physics\hspace*{0.2cm} conference\hspace*{0.2cm} reward\hspace*{0.2cm} ratio} = 
{{\rm \#\hspace*{0.2cm} allocated\hspace*{0.2cm} physics\hspace*{0.2cm} conference\hspace*{0.2cm} presentations}\over
{\rm total\hspace*{0.2cm} \#\hspace*{0.2cm} of\hspace*{0.2cm} physics\hspace*{0.2cm} internal\hspace*{0.2cm} papers\hspace*{0.2cm} produced+1}}.
\end{eqnarray}

Table~\ref{tab:comp} displays the averages of the total and physics conference reward ratios for the full sample of males and females.

We can see from the results of the table that there is a dramatic and significant gender bias in the allocation of
conference presentations; on average males have a conference reward ratio around 3 times that of females.
The physics conference reward ratio is also gender biased, with males receiving over twice the physics
conference presentations per physics internal paper than females.\footnote{If we take productivity out of the picture
and simply look at the number of conference allocations per year, males are allocated over 50\% more physics conference presentations
per year than females. Thus there is still a significant gender-dependent disparity in conference
allocations even without taking the high productivity
of the females into account.}

We note that the probability that a female moves on to a faculty position is significantly correlated
to her physics reward ratio.
It thus appears that the gender-bias in the allocation of conference presentations
appears to detrimentally impact the ability of females to move on to faculty positions.
As we will see later when we build a linear regression model to predict the career advancement of females, we find
that it is indeed the case that the dearth of physics conference presentations detrimentally
impacts the academic career advancement prospects of females.

\subsection{Researcher Socialization}
Most Run~II~Dzero internal papers are produced by multiple co-authors,
and just like the refereed papers produced by the experiment 
there are no first authors on these
internal papers; the co-authors are simply listed alphabetically.
There are literally hundreds of collaborators on the experiment, and
a measure of how socialized each researcher is within the collaboration
is the number of people with whom they typically co-author papers.  It is not
unusual for some internal papers to have in excess of 20 to 30 authors, particularly
papers that describe high-profile physics analyses.

The effective number of internal papers a research produces
is defined as
\begin{eqnarray}
{\rm effective \hspace*{0.2cm}\#\hspace*{0.2cm} papers} 
= \sum_{i=1}^{\rm \#\hspace*{0.2cm} of\hspace*{0.2cm} papers} 
{{1}\over{\rm \#\hspace*{0.2cm} co-authors
\hspace*{0.2cm} on}\hspace*{0.2cm} i^{\rm th}\hspace*{0.2cm} {\rm paper}}.
\end{eqnarray}
The socialization coefficient of the researcher is defined as
\begin{eqnarray}
{\rm socialization\hspace*{0.2cm} coefficient} 
= 1 - {{\rm effective\hspace*{0.2cm} \#\hspace*{0.2cm} papers}\over
{\rm total\hspace*{0.2cm} \#\hspace*{0.2cm} papers}}
\end{eqnarray}
A socialization coefficient
near 0 indicates that the researcher normally publishes alone, whereas a
coefficient near 1 indicates the researcher normally collaborates with large groups of people. 
In a similar fashion we can use the effective and total number of internal physics papers
produced by each researcher to define a physics socialization coefficient.
And similarly we can define a service work socialization coefficient.

Table \ref{tab:comp} shows the mean values
of the total, physics, and service work socialization coefficients for females and 
males.

We note from the table that there do not appear to be significant gender-dependent differences in
service work socialization, and that service work socialization is not correlated with becoming a professor.
In fact, male or female, become a professor or not, the average service work socialization is
consistently about $0.40$.

However, we see that for both males and females a high degree of physics socialization is significantly
correlated with becoming a professor, and that on average females appear to have a significantly higher physics
socialization coefficient than males.
In fact, females who become physics professors have almost exactly twice the average physics socialization
of females who do not.
As we will see later when we build a linear regression model to predict the career advancement of females, we find
that it is indeed the case that a low physics socialization coefficient detrimentally     
impacts the academic career advancement prospects of females.

Service work and physics socialization are not significantly correlated for either gender, leading us to conclude
that personal preference for working in small or large groups (which would apply to both physics and
service work) is not apparent.

%

\section{Modeling the probability of academic career advancement}

For males and females we develop separate binomial regression models that predict the probability of becoming a professor as
a function of the linear combination of physics and service work socialization, physics and service work productivity,
total and physics conference rewards, and the fraction of productivity devoted to service work.  
To reduce multicollinearity between the variables used, we normalize each of them using the mean and standard error
of the male data.\footnote{We normalize even the female data using the means and standard errors of the male data
such that we can directly compare the coefficients in the linear models of the males and females.}
We then trim each model using Akaike's Information Criterion (AIC) 
(see reference Kutner {\it et al.} (2005))
to obtain the most parsimonious
model that describes the data with minimal residual variance.

For the males, the best model is a linear combination of physics and service work productivity, physics
and service work socialization coefficients, and the fraction of productivity devoted to service work.
All the coefficients in the model corresponding to these terms are significant to $p\le2\%$.
This model correctly predicts the career outcome of 40 out of the 48 males, and has $R^2=0.51$.

The best model for females is a linear combination of their physics conference reward ratio and
their physics socialization coefficient. The coefficients in the model for
these two terms are both significant to $p\le0.10$. This model correctly predicts the career
outcome of 8 out of the 9 females, and has $R^2=0.59$.

The variables that best predict academic career advancement are gender dependent;
the reasons for this are primarily based on the differing productivity distributions for the
males and females; as noted before, exactly half the 48 males
have lower productivity than the least productive female.  The females are generally all highly
productive and the spread in their productivity is small. In contrast,
there is a wide spread in the productivity of the males.  Because the females
in general all have similar productivity, productivity clearly cannot play a role in deciding which
females move on to faculty positions.  However, for the males, it potentially can (and does)
have an impact on career advancement.  In fact, the physics productivity of males has a significantly
larger impact on the career outcome of the male researchers than any other variable included in the male model.
The coefficient for service work productivity in the male model is
much smaller in magnitude (and is also negative), meaning that service work productivity is much less
important compared to physics productivity to the career advancement of males.
If we add the fraction of productivity associated with service work to the model for the females
we find that the coefficient for that variable is also negative in the model,
indicating that service work appears to act as an impediment to the career advancement prospects
of females as well.

Also note that the best model for the males does not include a
conference
reward ratio, but the model for the females does.  
This is perhaps because the males have conference reward ratios 2 to 3
times that of the females, thus 
it is possible that the positive exposure to potential future employers reaches an
asymptotic maximum after a few conference presentations are allocated to a postdoctoral researcher, 
and after that the allocation of any extra conferences
does not provide any additional meaningful exposure.  The number of conferences allocated to
each female is on average so small that every conference appears to provide critical exposure
to potential future employers. 

One wonders how many more females could become faculty members
if the allocation of physics conference presentation was gender-blind,
and based purely on physics productivity.
To get some idea of this, we look at the physics productivity of each
female, and then find the male that has physics productivity closest
to this.  We then replace the conference reward ratio of the female 
with that of this male.
We then use our model for our females to predict the career outcomes
of these 9 ``non-discriminated-against'' theoretical females 
and find that 6 out of 9 are predicted to move on
to faculty positions (compared to 4 out of 9 in reality).

The coefficients corresponding to the physics socialization term in the male and female
models are statistically indistinguishable, and both are positive.
Thus physics socialization coefficients close to 1 correspond to significantly higher probability
of becoming a professor in both models.
For females, physics socialization is almost as critical to academic career advancement
as physics conference rewards, whereas physics socialization plays only a minor role
in deciding which males move on to faculty positions because of the very strong influence of
physics productivity in the male model.

\section{Discussion}
Based on a study of the productivity, conference presentation history,
and career paths of 57 former postdoctoral researchers on the Run~II~Dzero
experiment, 
we find quantitative examples of significant gender discrimination. 

We find that females were allotted 40\% more service work than males, 
and that the chances of this occurring in the absence of gender bias are 
less than 1\%.  This observation that females are significantly more often
shunted into service work roles 
echoes the results of  
a study performed 27 years ago by Mary Gaillard (1980) on the status of of female physicists
at CERN, a very large European particle physics laboratory.
Particle physics has not progressed very far in this respect in the last three decades.

We also find that females were significantly more productive
than their male peers in both physics and service work, 
yet were awarded significantly fewer conference
presentations; all 9 females in our sample were more productive than 24 out 
of the 48 males, yet the females had to be on average 3 times more productive than their
male peers in order to be awarded a conference presentation. The
chances of this occurring in the absence of gender bias are less than 1\%.
This result is in remarkable concordance with the research of Wenneras and Wold, who
found that females in their study had to be on average $2.5$ times more productive
than their male peers in order to receive a postdoctoral fellowship.

We note that this
dearth of allocated conference presentations 
appears to hinder the ability of otherwise highly qualified females
to become faculty members.  
Conference presentations give young physicists much needed exposure to future
employers, and state to the world outside the experiment that a particular researcher
is considered to be a productive member of the experiment.
It is not surprising then that
a dearth of physics conference presentations appears to detrimentally impact the
ability of a female to climb the academic career ladder.
In fact, our analysis finds that 
the physics conference reward ratio is the most
influential factor that determines which females go on to faculty positions, and
which do not.  Unlike socialization and productivity, this conference reward ratio
is completely out of the control of the researcher, and is thus an effective
gate-keeping mechanism that can be used by the upper administration of the
experiment to influence or impede the academic career advancement of females.

If the experiment allocated physics conference presentations based on physics productivity
rather than gender, we predict that around $50\%$ more females in our cohort would have moved on to faculty
positions.
It must be stressed that just because roughly equal fractions
of males and females in our cohort moved on to faculty positions does not mean that gender
equity is evident;  the females in our cohort have worked significantly harder than
their male peers to achieve this ``equity'' in academic career advancement, and yet some highly
competent female physicists nevertheless appear to be prevented from moving on to faculty positions
because of the conference allocation gate-keeping mechanism.

This study finds that the only other significant factor in female academic career advancement
is physics socialization;
some females are able to significantly socialize themselves into the collaboration,
such that they are put on author lists of physics papers with many authors, and
this high degree of socialization is strongly associated with their ability to
move on to a faculty position.
The physics socialization coefficient
of these select females is significantly greater than that of the average male, and
is nearly twice that of females who do not move on to faculty positions.
Again, this result is in concordance with the research of Wenneras and Wold.

It should be noted that the
service work and physics socialization coefficients
are not significantly correlated for either gender, leading us to conclude
that personal preference for working in small or large groups (which would apply to both physics and
service work) is not an apparent pattern in our data.
Instead, it is quite possible that
the high degree of physics socialization of some
females reflects the 
presence of a senior mentor who is actively engaged in promoting the career success
of their prot\'{e}g\'{e}e by ensuring that they work on high-profile physics analyses and
are well networked within the social fabric of the experiment.
Indeed, previous studies performed by Corcoran and Clark (1986) and
Cameron and Blackburn (1981) have shown a significant relationship between academic career
success and academic network
involvement, combined with sponsorship by senior faculty.
It is also possible that
some females perform work in large collaborative groups, but without the
presence of a strong mentor to ensure that their work
receives due credit, their contribution to the group
effort is subsequently overlooked on author lists of the internal 
papers associated with that
work;
previous studies and literature reviews performed by Bellas (1999) have shown that the gender-biased practice 
of not giving credit where credit has been earned
is common in academia due to differences
in how academics view male and female self-promotion and demands for credit for work performed.
It is an unfortunate limitation of this study that we do not have anecdotal
evidence from the sample of females to prove or disprove these specific
hypotheses for this cohort of data.  

Another limitation of this 
study is that it is specific to particle physics, and
is also specific to the Run~II~Dzero experiment; 
it would be interesting to see equivalent studies
performed with data obtained from other experiments with similar administrative infrastructures.
It would also be interesting to see such studies performed in other
sub-fields of physics.


\section{Summary}
This study 
follows the career paths of a full cohort of 57 postdoctoral
researchers who all worked under the same administrative infrastructure
during the same period of time.
The study determines the factors that appear to have the most influence
on the career paths of the members of the cohort, and whether or not
those factors are gender-dependent and/or within the control of the researcher.  
We find that the strongest influence
for the males is productivity, which is within the control of the researcher,
but that the
most significant influence on career path for females is conference allocations, 
which unfortunately are not in the control of the researcher.  

We also find that females were significantly more productive
than their male peers in both physics and service work, 
yet were awarded significantly fewer conference
presentations;  
the females in our cohort had to be on average 3 times more productive than their
male peers in order to be awarded a conference presentation. Our study predicts that
if conference presentations were allocated by the
administration of the experiment in a gender-blind fashion, we would 
expect that around 50\% more of the females in our cohort would have moved on to
faculty positions.
The gender-biased allocation of conference presentations to collaborators on the
experiment appears to be an effective gate-keeping mechanism that chooses which
females can move on to faculty positions and which cannot.

Gate-keeping mechanisms such as this can be addressed via enforcement of
Title~IX because this law unambiguously applies to people conducting research at any
federally funded U.S. national laboratory 
(whether they are employed by the laboratory or not)\footnote{The 
laboratory is obligated under Title IX to investigate and resolve any complaints
of gender discrimination perpetrated by the administration of the laboratory or
the administration of the 
experiments based at the laboratory.  Title~IX is unique among federal anti-discrimination statutes 
in that it protects not just employees of federally funded educational institutes or research laboratories,
but also anyone who performs research at such places as a visitor from another institute.}.
It is interesting to note that this conclusion is in concordance
with that of the 2004 report produced by the U.S. Government Accountability Office, USGAO (2004); the 
report examined the reasons behind the 
slow rise of the fraction of women participating in the sciences, and
called for greater Title IX compliance in federally funded research activities to address the problem.

\section{Acknowledgements}
The author wishes to thank Dr. Suzanne Franks, Dr. Dale Pitman, Professor Stephanie Goodwin of Purdue,
and Professor Mary Gaillard of UC Berkeley
for insightful discussions related to this work.

\section*{Appendix}
Some readers of this article may be wondering if the administration of Fermilab has
been made aware of this study; they have.  In late summer 2006 the author presented the preliminary results
of this study in a formal complaint  
to the Fermilab Equity
Office on behalf of all the female postdoctoral researchers collaborating on the Run~II~Dzero experiment.
The laboratory disdained to investigate the complaint despite the fact that many women were affected, and the fact that
it was likely the most thoroughly statistically well-founded complaint that office had ever received.

The author has since complained to the Department of Energy Office of Civil Rights (DoE OCR) regarding Title~IX
non-compliance at Fermilab (which receives federal grant
funding through the DoE).  It is the responsibility of the DoE OCR
to enforce Title~IX compliance in their federally funded research activities.  In the complaint, the author
pointed out that not only did Fermilab disdain to investigate the complaint, but also did not
meet even the minimum standards of Title~IX compliance (by, for instance, failing to publicly post the name
of the Title IX complaint co-ordinator on site, along with information about Title~IX and
instructions on how to complain about discrimination or harassment under that law).
The DoE OCR is currently investigating the complaint.


\newpage
\section*{References}

{\small
{\bf Bellas,M. (1999)} {\it Emotional Labor in Academia: The Case of Professors},\\ 
\underline{The Annals of the American Academy of Political and Social Science}, Volume 561, pp 96-111.\\[3mm]
{\bf Cameron,S. and Blackburn,R. (1981)} {\it Sponsorship and Academic Career Success},\\
\underline{Journal of Higher Education}, Volume 52, Number 4, pp 369-77.\\[3mm]
{\bf Coberly,W.A. and Lewis,T.O. (1972)}  
{\it A Note on a One-Sided Kolmogorov-Smirnov Test of Fit for Discrete Distribution Functions}, 
\underline{Annals of the Institute of Statistical Mathematics}, Volume 24, Number 1, pp 183-187.\\[3mm]
{\bf Corcoran,M. and Clark,S. (1986)} {\it Perspectives on the Professional Socialization of Women Faculty: 
A Case of Accumulative Disadvantage?}, \underline{Research in Higher Education}, Volume 57, Number 1, pp 20-43.\\[3mm]
{\bf Epstein,B. (1954)} 
{\it Tables for the Distribution of the Number of Exceedances}, \\
\underline{The Annals of Mathematical Statistics}, Volume 25, Number 4 , pp 762-768.\\[3mm]
{\bf Fisher,R.A. (1948)} {\it Combining Independent Tests of Significance}, 
\underline{American Statistician}, Volume~2, Issue 5, pp 30.\\[3mm]
{\bf Gaillard,M.K. (1980)} {\it Report On Women In Scientific Careers At Cern}, \underline{CERN/DG-11, Mar 1980}.\\[3mm]
{\bf Kutner,M. {\it et al.} (2005)} {\it Applied Linear Statistical Models}, McGraw-Hill.\\[3mm]
{{\bf Nelson,D. (2005)} {\it A National Analysis of Diversity in Science and Engineering Faculties at Research Universities},
(available at {\tt http://cheminfo.chem.ou.edu/faculty/djn/djn.html})
}\\[3mm]
{\bf Rolison,D. (2003)}  
{\it Can Title IX Do for Women In Science and Engineering What It Has Done for Women In Sports?}, 
\underline{APS News}, Volume 12.\\[3mm]
{\bf U.S. Government Accountability Office Report (2004)} {\it Women's Participation
in the Sciences Has Increased, but Agencies Need to Do More to Ensure Compliance with Title IX}, \underline{GAO-04-639}.\\[3mm]
{\bf Wenneras,C. and Wold,A. (1997)}
{\it Nepotism and Sexism in Peer-Review}, 
\underline{Nature} 337, pp 341.\\[3mm]
}

\clearpage
\begin{sidewaystable}
\begin{center}
\begin{tabular}{|l|c|c|c|}
\hline
                          & all     & all   & one-side    \\
                          & females & males & significance test        \\
                          & (n=9)   & (n=48)& probability \\
\hline
\# of internal papers/year                      & $1.70\pm0.39$ $(+0.17)$         & $1.38\pm0.17$ $({\bf +0.35})$ & $P_{\rm females>males}=0.98$ \\[2mm]
\# of physics internal papers/year              & $0.78\pm0.22$ $(+0.31)$         & $0.72\pm0.10$ $({\bf +0.42})$ & $P_{\rm females>males}=0.96$ \\[2mm]
\# of service work internal papers/year         & $0.92\pm0.21$ $(-0.01)$         & $0.66\pm0.09$ $(+0.11)$             & $P_{\rm females>males}=0.99$ \\[2mm]
fraction of internal papers                     & $0.66\pm0.10$ $({\bf -0.60})$ & $0.45\pm0.04$ $(+0.02)$             & $P_{\rm females>males}=0.95$ \\
devoted to service work & & & \\[2mm]
\hline
conference ``reward ratio''                         & $0.19\pm0.07$ $(+0.41)$         & $0.56\pm0.03$ $(-0.02)$             & $P_{\rm females<males}=0.99$ \\[2mm]
physics conference ``reward ratio''                 & $0.34\pm0.12$ $({\bf +0.46})$     & $0.78\pm0.05$ $(-0.05)$             & $P_{\rm females<males}=0.95$ \\[2mm]
\hline
socialization coefficient                           & $0.59\pm0.11$ $({\bf +0.62})$ & $0.57\pm0.05$ $(+0.15)$             & $P_{\rm females>males}=0.73$ \\[2mm]
physics socialization coefficient                   & $0.58\pm0.15$ $({\bf +0.60})$ & $0.47\pm0.06$ $({\bf +0.38})$ & $P_{\rm females>males}=0.96$ \\[2mm]
service work socialization coefficient              & $0.42\pm0.11$ $(+0.36)$         & $0.41\pm0.05$ $(+0.09)$             & $P_{\rm females>males}=0.86$ \\[2mm]
\hline
\end{tabular}
\end{center}
\vspace*{-0.5cm}
\caption[foo]{
\label{tab:comp}
Averages of productivity measures, conference rewards, and socialization coefficients for males and females.
Also shown is the standard error on each average.
In brackets by each average is the correlation of each variable to the probability that a female or male becomes a faculty member.
Correlations shown in bold face are significantly different from zero to $p\le0.10$.
The last two columns contain the one-sided significance test probabilities
comparing the distribution of the sample of females for a particular variable to the corresponding variable for the sample of males.
A lower-side $P_{\rm females<males}$ (upper-side $P_{\rm females>males}$) probability that is close to 1
indicates that the variable for the females is significantly smaller (larger) than that of the typical male.
}
\end{sidewaystable}

\end{document}